\documentclass[10pt, conference, compsocconf]{IEEEtran}
\usepackage{cite}
\usepackage{url}
\usepackage{amssymb}

\newcommand{\tb}{\textbf}

\newcommand{\ts}{\textsf}
\newcommand{\ttt}{\texttt}

\newcommand{\mi}{\mathit}

\begin{document}

\title{A State Space Tool for Models Expressed In C++\\
(tool paper)}

\author{\IEEEauthorblockN{Antti Valmari}
\IEEEauthorblockA{Department of Mathematics\\
Tampere University of Technology\\
Tampere, Finland\\
Email: Antti.Valmari@tut.fi}}

\maketitle

\begin{abstract}
This publication introduces A State Space Exploration Tool that is based on
representing the model under verification as a piece of C++ code that obeys
certain conventions.
Its name is ASSET.
Model checking takes place by compiling the model and the tool together, and
executing the result.
This approach facilitates very fast execution of the transitions of the model.
On the other hand, the use of stubborn sets and symmetries requires that
either the modeller or a preprocessor tool analyses the model at a syntactic
level and expresses stubborn set obligation rules and the symmetry mapping as
suitable C++ functions.
The tool supports the detection of illegal deadlocks, safety errors, and may
progress errors.
It also partially supports the detection of must progress errors.
\end{abstract}


\section{Introduction}\label{S:intro}

This publication discusses A State Space Exploration Tool called ASSET.
It is written in C++.
The model under verification is represented as a piece of C++ code that uses
the pre-defined type \ttt{state\_var} and implements certain functions such as
\ttt{fire\_transition}, \ttt{print\_state}, and \ttt{check\_state}.
The model is checked by copying it to the file \ttt{asset.model} and then
compiling and executing the file \ttt{asset.cc}.
The latter \ttt{\#include}s the former.
So the model is compiled into machine code instead of being simulated by
ASSET.

This approach leads to very fast execution of the transitions of the model.
It also gives great flexibility, because most features of C++ are available
for writing the model.
On the other hand, it sometimes leads to unnatural-looking models.
Furthermore, as such, ASSET cannot perform any syntactic analysis on the
model.

Some advanced methods such as stubborn sets and symmetries require such
analysis.
The modeller may perform the necessary analysis manually and represent the
result as certain C++ functions.
Alternatively, there could be a preprocessor tool that inputs some
user-friendly modelling language.
The preprocessor reads the model, analyses it at the syntactic level, and
writes it and the necessary additional functions in the form suitable for
ASSET.
The present author hopes that researchers will find ASSET useful and implement
preprocessors from their favourite languages.
ASSET is available free of charge for scientific and academic use at
\ts{http://www.cs.tut.fi/$\sim$ava/ASSET/}.

Throughout this publication, a \emph{demand-driven token ring} is used as an
example.
It consists of $n$ customers and $n$ servers.
The customers may request and be granted access to a critical section, return
from it to the initial local state, and, for reasons discussed
in~\cite{Val15}, terminate for good.
One \emph{token} circulates in the system.
Only the server that has the token may grant access to its customer.
To avoid unnecessary work, the token is not circulated when no customer is
requesting access.
When necessary, wait information is propagated in the opposite direction to
the token.

The model of the example system is introduced in Section~\ref{S:dtknring}.
Section~\ref{S:check} discusses the features that ASSET offers for specifying
correctness properties.
Sections~\ref{S:stubborn} and~\ref{S:symmetry} focus on the use of the
stubborn set and symmetry methods in ASSET.
Results of the experiments with the example system are collected in
Table~\ref{T:results}.
The times are in seconds and do not include the compilation.


\section{The Demand-Driven Token Ring}\label{S:dtknring}

\begin{figure*}\small
\begin{verbatim}
#ifdef size_par
const unsigned n = size_par;  // number of customers and servers from compilation command
#else
const unsigned n = 6;         // default number of customers and servers
#endif

state_var
  C[n] = 2, // state of customer i: 0 = idle, 1 = requested, 2 = critical, 3 = terminated
  S[n] = 2, // state of server i: 0 = idle, 1 = waiting for token, 2 = waiting for customer
  T[n] = 1; // true <==> server i has token

#ifdef symm_must
state_var c0now;  // the current index of the original customer 0
#else
unsigned const c0now = 0;
#endif

const char Cchr[] = { '-', 'R', 'C', ' ' }, Schr[] = { 'i', 'w', 't' };
void print_state(){
  for( unsigned i = 0; i < n; ++i ){
    std::cout << Cchr[C[i]] << Schr[S[i]];
    if( T[i] ){ std::cout << '*'; }else{ std::cout << ' '; }
  }
  std::cout << '\n';
}

unsigned nr_transitions(){ T[1] = true; return 3*n; }

inline unsigned next( unsigned i ){ return (i+1) % n; }
inline unsigned prev( unsigned i ){ return (i+n-1) % n; }

bool fire_transition( unsigned i ){

  /* Servers */
  if( i >= 2*n ){
    i -= 2*n;
    #define goto(x){ S[i] = x; return true; }
    switch( S[i] ){
    case 0: if( C[i] == 1 || ( S[next(i)] == 1 && !T[next(i)] )){ goto(1) }
            return false;
    case 1: if( !T[i] ){ return false; }
            if( C[i] == 1 ){ C[i] = 2; goto(2) }
            if( S[next(i)] == 1 ){ T[i] = false; T[next(i)] = true; goto(0) }
            return false;
    case 2: if( C[i] == 2 ){ return false; }
            T[i] = false; T[next(i)] = true; goto(0)
    default: err_msg = "Illegal local state"; return false;
    }
  }

  /* Customers */
  #undef goto
  #define goto(x){ C[i] = x; return true; }
  if( i >= n ){   // termination transition
    i -= n;
    if( C[i] == 0 ){ goto(3) }else{ return false; }
  }
  if( C[i] == 0 ){ goto(1) }  // request access
  if( C[i] == 2 ){ goto(0) }  // leave critical
  return false;
}
\end{verbatim}
\caption{Model of the demand-driven token ring.}\label{F:Dring}
\end{figure*}

\begin{table*}
\begin{tabular}{@{}r|rrr|rrr|rrr|rrr@{}}
    & \multicolumn{3}{c|}{plain} & \multicolumn{3}{c|}{stubborn sets}
    & \multicolumn{3}{c|}{symmetries} & \multicolumn{3}{c}{both} \\
$n$ & states & edges & time & states & edges & time
    & states & edges & time & states & edges & time\\
\hline
 2 &           68 &           140 & 0.0 &           44 &            60 & 0.0
   &           34 &            70 & 0.0 &           22 &            30 & 0.0\\
 3 &          468 &        1\,350 & 0.0 &          219 &           327 & 0.0
   &          156 &           450 & 0.0 &           73 &           109 & 0.0\\
 4 &       2\,928 &       10\,880 & 0.0 &          920 &        1\,432 & 0.0
   &          732 &        2\,720 & 0.0 &          230 &           358 & 0.0\\
 5 &      17\,280 &       78\,600 & 0.1 &       3\,505 &        5\,625 & 0.0
   &       3\,456 &       15\,720 & 0.0 &          701 &        1\,125 & 0.0\\
 6 &      98\,064 &      527\,760 & 0.2 &      12\,540 &       20\,772 & 0.1
   &      16\,344 &       87\,960 & 0.1 &       2\,090 &        3\,462 & 0.0\\
 7 &     541\,296 &   3\,364\,200 & 0.8 &      43\,015 &       73\,899 & 0.2
   &      77\,328 &      480\,600 & 0.4 &       6\,145 &       10\,557 & 0.1\\
 8 &  2\,927\,232 &  20\,632\,320 & 4.5 &     143\,408 &      256\,880 & 0.4
   &     365\,904 &   2\,579\,040 & 1.6 &      17\,926 &       32\,110 & 0.2\\
 9 & 15\,583\,104 & 122\,821\,920 &30.0 &     469\,053 &      879\,885 & 1.4
   &  1\,731\,456 &  13\,646\,880 &10.0 &      52\,117 &       97\,765 & 0.3\\
10 & 81\,933\,120 & 714\,052\,800 & 262 &  1\,514\,900 &   2\,984\,860 & 4.6
   &  8\,193\,312 &  71\,405\,280 &59.5 &     151\,490 &      298\,486 & 0.9\\
11 &           -- &            -- & 341 &  4\,852\,771 &  10\,057\,839 &16.3
   & 38\,771\,136 & 370\,202\,400 & 339 &     441\,161 &      914\,349 & 2.6\\
12 &              &               &     & 15\,464\,040 &  33\,719\,400 &60.1
   &           -- &            -- &1039 &  1\,288\,670 &   2\,809\,950 & 9.1\\
13 &              &               &     &           .. &            .. &65.0
   &              &               &     &  3\,777\,949 &   8\,659\,221 &30.0\\
14 &              &               &     &              &               &
   &              &               &     & 11\,116\,762 &  26\,741\,542 &96.1\\
15 &              &               &     &              &               &
   &              &               &     & 32\,826\,001 &  82\,708\,765 &353\\
16 &&&&&&&&&& .. & .. & 131\\
\end{tabular}
\caption{Results on the demand-driven token ring. ``--'' denotes that $10^8$
states was exceeded. ``..'' indicates memory overflow.}\label{T:results}
\end{table*}

Figure~\ref{F:Dring} shows the model of the example system.
The \ttt{\#ifdef} \ttt{size\_par} structure makes it possible to specify the
number of customers and servers via an option that is given to the C++
compiler.
Also many features of ASSET can be controlled in a similar manner.
For instance, with the Gnu C++ compiler, the options \ttt{-Dstubborn}
\ttt{-Dsize\_par=13} \ttt{-Dstop\_cnt=100000000} command ASSET to use stubborn
sets, set $n=13$, and stop the construction of the state space when $10^8$
states are exceeded.

The initialization \ttt{C[n]} \ttt{=} \ttt{2} does not specify that the
initial local state of each customer is 2 (the critical section) but that two
bits are used for representing the local state of each customer.
The value of each state variable is an unsigned integer in the range $0,
\ldots, 2^b - 1$, where $b$ is the number of bits.
The initial value is $0$.
The default value of $b$ is 8.

Internally, ASSET represents the state of the model as a sequence of unsigned
integers.
To save memory, ASSET packs many state variables into the same unsigned
integer when possible.
If the most recently employed unsigned integer has at least as many unused
bits as the next state variable needs, then ASSET puts the state variable
there.
This implies that the order in which the state variables are declared may
affect the amount of memory that ASSET uses per state.
For instance, assuming 64-bit unsigned integers,
\begin{verbatim}
      state_var x(3), A[16]=4, y=1;
\end{verbatim}
consumes three unsigned integers, while
\begin{verbatim}
      state_var A[16]=4, x(3), y=1;
\end{verbatim}
consumes only two.

When ASSET has detected an error, it prints a counterexample in the form of a
sequence of states.
For this purpose, it needs a \ttt{print\_state} function.
To improve the readability of the counterexamples, the function in
Figure~\ref{F:Dring} uses character encodings for local states.

The function \ttt{nr\_transitions} tells ASSET how many \emph{transitions} the
model contains.
The most common case is that a transition models one or more atomic operations
of the system.
Transitions in ASSET are deterministic.
This implies that nondeterministic operations such as tossing a coin must be
modelled by more than one transition.
Other than that, ASSET does not restrict the grouping of atomic operations to
transitions.

Each server of the example system has one transition.
In its initial state, a customer makes a nondeterministic choice between
requesting access and terminating for good.
For this reason, two transitions are used to model each customer.

The function \ttt{nr\_transitions} may also be used for implementing whatever
operations are necessary before starting the construction of the state space.
In Figure~\ref{F:Dring}, it is used for giving the token to customer 1.

The transitions of the model are numbered starting from~0.
The function \ttt{fire\_transition($t$)} returns \ttt{true} or \ttt{false} to
indicate whether transition number $t$ is enabled.
If $t$ is enabled, then \ttt{fire\_transition} changes the state according to
the occurrence of $t$.
If $t$ is disabled, then \ttt{fire\_transition} must not change the state.
This rule makes it possible for ASSET to try the next transition without
having to upload the state again.

To improve readability, Figure~\ref{F:Dring} introduces two versions of a
\ttt{goto(x)} macro.
They model the server and the customer going to local state \ttt{x} and
indicate that the transition was enabled.

If $0 \leq t < n$, transition $t$ models customer $t$ going either from local
state 0 to local state 1 (that is, requesting access) or from local state 2 to
local state 0 (leaving the critical section).
If $n \leq t < 2n$, transition $t$ models customer $t-n$ going from local
state 0 to local state 3 (that is, terminating).

Finally, the transitions $2n \leq t < 3n$ model server $t-2n$.
It waits in local state 0 until its customer requests access or the next
server needs the token.
For the reason discussed in Section~\ref{S:stubborn}, it tests that the next
server does not already have the token.
Then it waits in local state~1 until it has the token.
If its customer has requested access, it moves the customer to the critical
section and goes to local state 2.
Otherwise, if the next server needs the token, server $t-2n$ gives it to it.
Otherwise, server $t-2n$ continues waiting.

In local state 2, server $t-2n$ waits until its customer has left the
critical section.
Then it gives the token to the next server and returns to the idle state.
As a consequence, its customer cannot get access again before the token has
circulated through the ring and the other customers have had the chance to get
access.

\section{The Checking Features}\label{S:check}

\begin{figure*}\small
\begin{verbatim}
/* Check that at most one customer is in critical section at any time. */
#define chk_state
const char *check_state(){
  unsigned cnt = 0;
  for( unsigned i = 0; i < n; ++i ){ if( C[i] == 2 ){ ++cnt; } }
  if( cnt >= 2 ){ return "Mutual exclusion violated"; }
  return 0;
}

/* Check that every customer has stopped. */
#define chk_deadlock
const char *check_deadlock(){
  for( unsigned i = 0; i < n; ++i ){
    if( C[i] != 3 ){ return "Customer not terminated"; }
  }
  return 0;
}

/* Check that the original customer 0 eventually gets access if it wants to. */
//#define chk_must_progress
bool is_must_progress(){ return C[c0now] != 1; }
\end{verbatim}
\caption{The check functions of the model of the demand-driven token
ring.}\label{F:check}
\end{figure*}

Figure~\ref{F:check} shows the checking functions used in the experiments of
this publication.
Each of them can be switched off by commenting out the corresponding
\ttt{\#define}, without having to comment out the function as a whole.
This is handy for experimenting.
(It would have been nice to use the same word in the \ttt{\#define} and as the
name of the function, but C++ does not allow that.)
More flexibility comes from the fact that if \ttt{xxx} has not been switched
on in the model with \ttt{\#define} \ttt{xxx}, then it can be switched on at
compile time with a compiler option.

ASSET operates in stages.
In the first stage, it checks for safety errors and illegal deadlocks (if the
checking of them has been switched on).
It constructs the state space in breadth-first order, to minimize the length
of counterexamples.
ASSET calls \ttt{check\_state} each time when it has constructed a new state,
and \ttt{check\_deadlock} when it has tried to fire transitions in a state but
none was enabled.
If the state is not good, the function returns a character string.
ASSET prints an error message containing it and terminates.
That is, ASSET implements on-the-fly detection of safety and illegal deadlock
errors.
That the state is good is indicated by returning the null pointer \ttt{0}.

If ASSET did not detect any errors and if it has further checks to perform, it
constructs a data structure that contains the edges of the state space in the
reverse direction.
To do that, it goes through all states that it has found and fires the same
transitions again in them as it fired in the first stage.
In this way, only one unsigned integer per edge is needed.
Storing the edges during the first stage and sorting them afterwards would
have used two unsigned integers per edge.

Then, if \ttt{chk\_may\_progress} is on, ASSET checks the state space for
\emph{may progress errors} by performing a linear-time search along the
reversed edges.
A may progress error is a reachable state from which no terminal state and no
state accepted by \ttt{is\_may\_progress} is reachable.
May progress can be thought of as a less stringent alternative to linear-time
liveness that does not need fairness assumptions.
This feature has been discussed extensively in~\cite{Val15} and is not used in
the experiments of the present publication, so it will not be discussed
further here.

Next ASSET checks the state space for \emph{must progress errors}, if it has
been commanded to do so and it has not yet terminated because of another
error.
A must progress error is a cycle in the state space that does not contain any
state accepted by \ttt{is\_must\_progress}.
This is a restricted form of checking linear-time liveness.
For reasons discussed in the next two sections, the simultaneous use of this
feature with stubborn sets and symmetries is limited.
Therefore, it has been switched off in Table~\ref{T:results}.

Outside Table~\ref{T:results}, the \ttt{is\_must\_progress} function in
Figure~\ref{F:check} was switched on in some experiments.
ASSET found no errors in the model in Figure~\ref{F:Dring}.
Also a modified model was used where, when leaving local state 2, instead of
giving the token to the next server and going to local state~0, the server
goes to local state 1.
ASSET reported that this model has a cycle where a requesting customer does
not get access.
In it, another customer leaves the critical section, requests for access
again, and gets access again.

Finally, if the stubborn set method is used, safety or progress was checked,
and ASSET has not yet found any error, it checks that the state space is
\tb{AG} \tb{EF} terminating in the sense discussed in the next section.

The model may at any time assign a character string to \ttt{err\_msg}.
It causes ASSET to terminate and print an error message containing the string.
This feature is not intended for specifying correctness properties, but for
catching inconsistent situations within the model.
In Figure~\ref{F:Dring} it is used in the \ttt{default} branch of the
\ttt{switch} statement, to indicate that the modeller believes that the branch
is never entered.

In addition to the memory needed for the state itself, ASSET uses two or five
unsigned integers per state, depending on whether the verification task
involves graph search operations in the state space.
This explains the ``..'' entries in Table~\ref{T:results}.

\section{The Stubborn Set Method in ASSET}\label{S:stubborn}

\begin{figure*}\small
\begin{verbatim}
void next_stubborn( unsigned i ){

  if( i >= 2*n ){
    i -= 2*n;
    switch( S[i] ){
    case 0: if( C[i] == 1 || ( S[next(i)] == 1 && !T[next(i)] )){ return; }
            stb(i, next(i)+2*n); return;
    case 1: if( !T[i] ){ stb(prev(i)+2*n); return; }
            if( C[i] == 1 ){ return; }
            if( S[next(i)] == 1 ){ stb(i); return; }
            stb(i, next(i)+2*n); return;
    case 2: if( C[i] == 2 ){ stb(i); }
            return;
    default: return;
    }
  }

  if( i >= n ){ stb(i-n); return; }
  switch( C[i] ){
  case 0: stb(i+n, i+2*n); return;
  case 1: stb(i+2*n); return;
  case 2: stb_all(); return;
  default: return;
  }
}
\end{verbatim}
\caption{The stubborn set obligation rules of the model of the demand-driven
token ring.}\label{F:stubborn}
\end{figure*}

The implementation of the stubborn set method in ASSET is discussed
extensively in~\cite{Val15}.
Therefore, it is discussed here only briefly.

Only the basic strong stubborn set method is implemented.
However, theorems in~\cite{Val15} tell that if the model is \tb{AG} \tb{EF}
terminating --- that is, if from every reachable state, a terminal state is
reachable, then the basic strong stubborn set method preserves also safety and
certain progress properties.
Furthermore, whether the model is \tb{AG} \tb{EF} terminating can be checked
from the reduced state space.

To use the method, the function \ttt{next\_stubborn} must be provided.
It represents state-dependent rules of the form ``if this transition is in the
stubborn set, then also these transitions must be''.
Figure~\ref{F:stubborn} shows the rules used in the experiments reported in
Table~\ref{T:results}.

The present author did not at first realize the necessity of the part
\ttt{\&\&} \ttt{!T[next(i)]} in case 0 in Figure~\ref{F:Dring}.
When it was lacking, the plain and symmetry methods did not give any error
messages.
Indeed, mutual exclusion and eventual access are not violated.
However, thanks to the check that the model is \tb{AG} \tb{EF} terminating,
the stubborn set method gave the following when $n=2$.

{\small\begin{verbatim}
-i -i*
-i  i*
==========
Ri  i*
Rw  i*
Rw  w*
Rw* i
Rw* w
Ct* w
-t* w
-i  w*
----------
 i  w*
 w  w*
 w* i
 w* w
!!! State was reached from which termination
    is unreachable
67 states, 93 edges
\end{verbatim}}
\begin{figure*}\small
\begin{verbatim}
void symmetry_representative(){
  unsigned i = 0;
  while( !T[i] ){ ++i; }            // find the server with the token
  i = prev(i); if( !i ){ return; }  // terminate, if the state maps to itself
  unsigned A[n], j;
  for( j = 0; j < n; ++j ){ A[j] = C[(i+j) % n]; }
  for( j = 0; j < n; ++j ){ C[j] = A[j]; }
  for( j = 0; j < n; ++j ){ A[j] = S[(i+j) % n]; }
  for( j = 0; j < n; ++j ){ S[j] = A[j]; }
  for( j = 0; j < n; ++j ){ A[j] = T[(i+j) % n]; }
  for( j = 0; j < n; ++j ){ T[j] = A[j]; }
  #ifdef symm_must
  c0now = (c0now + n-i) % n;
  #endif
}
\end{verbatim}
\caption{The symmetry representative function of the demand-driven token
ring.}\label{F:symmetry}
\end{figure*}

In it, customer 0 visits the critical section and both customers terminate.
Because waiting information may be propagated from server $\mi{next}(i)$ to
server $i$ even if the former has the token, unnecessary waiting information
enters the ring.
Eventually the model runs in a cycle where unnecessary waiting information
circulates in one direction and, driven by it, the token circulates in the
opposite direction.
The cycle consists of the states below the ``\ttt{----------}'' mark.
Reaching a terminal state is impossible after the ``\ttt{==========}'' mark.
So the first erroneous state is \ttt{Ri} \ttt{i*}.
The length of the counterexample has not been minimized.

The stubborn set implementation in ASSET does not guarantee that must progress
errors are found.
If must progress is used with stubborn sets and ASSET finds no errors, then it
gives a warning that the pass verdict is unreliable.
With the modified model discussed in Section~\ref{S:check}, ASSET did find
the error with stubborn sets switched on.
With $n=8$, there were 2\,472\,336 states and 17\,539\,200 edges without and
163\,264 states and 293\,984 edges with stubborn sets.

\section{The Symmetry Method in ASSET}\label{S:symmetry}

Many systems contain similar components organized in a symmetric fashion.
Several authors have suggested exploiting the symmetry for reducing the size
of the state space, including~\cite{CFJ93,EmS93,Jen95}.

The implementation of the symmetry method in ASSET is very simple.
Unfortunately, as we will see, it leaves a lot of responsibility to the
modeller or preprocessor tool.

The modeller or preprocessor must provide the function
\ttt{symmetry\_representative}.
It must map each state to a symmetric state.
The more states are mapped to the same state, the better are the reduction
results.
Ideally, all states that are symmetric to each other are mapped to the same
state.
However, the method remains correct even if the function is not ideal in this
respect.

If the symmetry method has been switched on, ASSET calls
\ttt{symmetry\_representative} on the initial state and on each result of a
successful firing of a transition.
As a consequence, paths in the reduced state space may contain \emph{symmetry
swaps}, that is, the head state of an edge is not necessarily the real result
of firing the transition in question in the tail state of the edge.
Instead, it may be another state that is symmetric to the real result.

Figure~\ref{F:symmetry} shows the symmetry mapping used in the experiments of
this publication.
It first finds the server that has the token, and then rotates the ring so
that the found server becomes server 1.

The modeller or preprocessor must take the symmetry mapping into account when
formulating the checked properties.
Because \ttt{check\_state} and \ttt{check\_deadlock} analyse a single state,
it is not difficult to make them give the same reply on symmetric states.
The versions in Figure~\ref{F:check} do so.

It is more difficult with progress properties.
For instance, consider the eventual access property ``if customer 0 wants to
go to the critical section, it eventually gets there''.
When \ttt{symm\_must} is off, the \ttt{is\_must\_progress} function in
Figure~\ref{F:check} formulates the property in a manner that is appropriate
only when the symmetry method is not used.
Because the symmetry mapping in Figure~\ref{F:symmetry} always rotates the
system such that server 1 has the token, only customer~1 ever gets to the
critical section in the symmetry-reduced state space.
However, the rotation does not prevent customer 0 from trying to go to the
critical section.
What happens is that just when customer 0 is about to get to the critical
section, it becomes customer~1.
So ASSET incorrectly reports that customer 0 tried to go to the critical
section but never got there.

To solve this problem, the state variable \ttt{c0now} was added to the model.
It keeps track of the current number of the original customer 0.
The verification results became correct, but the reduction in the size of the
state space was lost entirely.
This is a problem of not just ASSET, but the symmetry method in general.

The counterexamples printed by ASSET may contain symmetry swaps.
However, in the experience of the present author, they have not harmed the
interpretation of counterexamples.
They may even be helpful.
When a counterexample contains many symmetric copies of the same theme,
symmetry swaps may reduce them to a single copy.

\section*{Acknowledgement}

I thank the anonymous reviewers for helpful comments.


\end{document}